\documentclass[aps,pre,twocolumn]{revtex4-1}
\usepackage{epsfig}

\newcommand{\be}{\begin{eqnarray}}
\newcommand{\ee}{\end{eqnarray}}
\newcommand{\no}{\nonumber}

\begin{document}
\title{Dynamical Singularities of Glassy Systems in a Quantum Quench}
\author{Tomoyuki Obuchi}
\affiliation{Cybermedia Center, Osaka University, Osaka 560-0043, Japan}
\affiliation{CNRS-Laboratoire de Physique The\'orique de l'ENS, 24 rue Lhomond, 75005 Paris, France}
\author{Kazutaka Takahashi}
\affiliation{Department of Physics, Tokyo Institute of Technology, 
Tokyo 152-8551, Japan}

\date{\today}

\begin{abstract}
We present a prototype of behavior of glassy systems 
driven by quantum dynamics in a quenching protocol by analyzing 
the random energy model in a transverse field. 
We calculate several types of dynamical quantum amplitude and 
find a freezing transition at some critical time. 
The behavior is understood by the partition-function zeros 
in the complex temperature plane.
We discuss the properties of the freezing phase as a dynamical chaotic phase, 
which are contrasted to those of the spin-glass phase in the static system.
\end{abstract}


\maketitle
\section{INTRODUCTION}

Phase transitions are one of the most fascinating topics 
in statistical physics.
After the long investigations, equilibrium phase transitions are clarified 
as singularities of free energies and become known to be treatable 
in the current framework of statistical mechanics~\cite{
Onsager:44,Yang:52,Fisher:64}. 
To analyze those singularities, sophisticated techniques and concepts, 
such as the renormalization group theory and the scaling hypothesis, 
have been founded and nowadays equilibrium phase transitions of 
classical systems are fairly well understood. 
On the other hand, phase transitions or dynamical singularities of 
quantum systems are far less understood, though rising interests 
in quantum computation require better comprehension of those phenomena. 
Recent experimental developments enable to observe time evolutions of 
quantum many-body systems 
in detail~\cite{Greiner:02,Kinoshita:06,Sadler:06}, 
which also demands better theoretical clarifications of 
dynamical behavior of quantum systems.

Recently, dynamical singularities of quantum systems in a typical 
experimental set up, the so-called quantum quench, 
are focused~\cite{Pollmann:10,Kolodrubetz:11,Heyl:12}. 
In the set up, a quantum system, whose Hamiltonian with 
a parameter $\Gamma$ is written as $\mathcal{H}(\Gamma)$, 
is prepared in a quantum state $|\psi_0\rangle$ dominated 
by $\mathcal{H}(\Gamma_{0})$, and the parameter is suddenly 
changed as $\Gamma_{0}\to \Gamma_{1}$ at $t=0$. 
After the quench, the wave function of the system 
at time $t\geq 0$ is given by
\be
 |\psi(t)\rangle=e^{-it \mathcal{H}(\Gamma_1)}|\psi_0\rangle.
\ee
Due to the sudden quench, the behavior of this wave function can be 
drastically different from the one in adiabatic process. 
It is actually reported that another energy scale different from 
the adiabatic one exists and dominates the system, 
which can lead to novel non-analyticity in dynamical 
behavior~\cite{Pollmann:10,Kolodrubetz:11,Heyl:12}. 
Interestingly, those dynamical singularities can be analyzed by 
the theory of zeros of equilibrium partition 
function~\cite{Yang:52,Fisher:64,Heyl:12}. 
This provides a possibility to analyze a wide range of dynamical 
singularities of quantum systems, 
with the current maturity of the theory of zeros.

Based on this possibility, in this paper, 
we provide a prototype of dynamical singularities of glassy systems. 
For this, we focus on the so-called 
random energy model (REM)~\cite{Derrida:80,Gross:84} 
in a transverse field (TREM) introducing quantum 
nature~\cite{Goldschmidt:90,Obuchi:07,Jorg:08,Takahashi:11}. 
Despite its simplicity, the REM is known to show many nontrivial 
properties of glassy systems and is treated as an ideal platform 
to study systems with many metastable 
states~\cite{Kirkpatrick:89,Bryngelson:87}, 
which naturally motivates us to firstly investigate the TREM 
to examine dynamical singularities of glassy systems.
In addition, the distribution of 
the partition-function zeros is well known 
in the REM~\cite{Derrida:91,Takahashi:11}, and 
we can expect that the present analysis reveals 
a clear relation between the dynamical singularities and zeros.

\section{MODEL}

The Hamiltonian of the TREM is defined as the limit of 
the fully-connected Ising spin-glasses (SGs) with $p$-body interactions 
$\mathcal{H}=H_{p}+H_{g}$ where
\be
 & & H_{p} = 
 -\sum_{i_1<\cdots<i_p}J_{i_1\cdots i_p}\sigma_{i_1}^z\cdots\sigma_{i_p}^z, 
  \label{eq:SG-Hamiltonian}
 \\
 & & H_g = -\Gamma\sum_{i=1}^N\sigma_i^x.
\ee
$\sigma^{x,z}$ are Pauli matrices, $N$ is the number of spins, and 
the interaction $J_{i_1\cdots i_{p}}$ is independent identically 
distributed (i.i.d.) from Gaussian 
with zero mean and variance $p!/2N^{p-1}$. 
The REM is produced by taking the limit 
$\lim_{p\to\infty}H_{p}= H_{\rm REM}$~\cite{Derrida:80}.
Energy levels of $H_{\rm REM}$ lose the correlations among each other 
in the limit $p\to \infty$, and thus energy levels of the REM 
Hamiltonian $H_{\rm REM}$ become i.i.d. Gaussian variables 
with zero mean and variance $N/2$. 
The TREM Hamiltonian is given by $\mathcal{H}_{\rm TREM}=H_{\rm REM}+H_{g}$, 
and hence in the $\sigma^{z}$ representation the diagonal elements are 
random Gaussian variables coming from $H_{\rm REM}$, 
and the off-diagonal ones arise from $H_{g}$. 

The TREM has three thermodynamic phases~\cite{Goldschmidt:90}. 
One is the classical paramagnetic (CP) phase which dominates 
the system for small $\Gamma$ and large temperature $T=1/\beta$, 
and another one is the SG phase appearing at low $T$ 
and small $\Gamma$. 
The last one is the quantum paramagnetic (QP) phase 
emerging for large $\Gamma$. 
The free energy densities are given by
\be
 f = \left\{\begin{array}{lc}
 -\frac{\beta}{4}-\frac{\ln 2}{\beta} & \mbox{for CP} \\
 -\sqrt{\ln 2} & \mbox{for SG} \\
 -\frac{1}{\beta}\ln (2\cosh \beta\Gamma) & \mbox{for QP}
 \end{array}\right..
\ee
First-order phase transitions from QP to CP or SG phases occur at 
critical values $\Gamma_{\rm c}(T)$. 
Clearly, the CP and SG phases do not depend on $\Gamma$, 
and also the QP phase is not influenced by $H_{\rm REM}$ 
since $f$ is completely identical to 
$-(1/N\beta)\ln {\rm Tr}\,\exp(-\beta H_{g})$. 
These imply the strong simplicity of the TREM. 
Actually, J\"{o}rg et al. showed that perturbations from $\Gamma=0$ 
and $\Gamma=\infty$ do not change the eigenfunctions 
in both two cases in the thermodynamic limit~\cite{Jorg:08}, 
which can be schematically written as
\be
 {\rm Tr}\,e^{-\beta\mathcal{H}_{\rm TREM}}\to \left\{
 \begin{array}{ll}
 {\rm Tr}\,e^{-\beta H_{\rm REM}} & \mbox{for}\ \Gamma<\Gamma_{\rm c}(T) \\
 {\rm Tr}\,e^{-\beta H_g}       & \mbox{for}\ \Gamma\geq\Gamma_{\rm c}(T)
 \end{array}	      
 \right.. \label{trem}
\ee
This property makes the following analysis extremely simple.

\section{OVERLAPS}

\subsection{Return amplitude}

Hereafter, we consider a quantum quench protocol from large to 
small $\Gamma$, and examine the dynamical singularities in the TREM. 
Following Ref.~\cite{Heyl:12}, 
we investigate the return amplitude 
$G(t)\equiv\langle\psi_0|\psi(t)\rangle=
\langle\psi_0|\exp(-it\mathcal{H})|\psi_0\rangle$, 
especially focusing on the ground state for large $\Gamma$. 
Namely, we put $|\psi_0\rangle=|{\rm QP}\rangle$ where 
$|{\rm QP}\rangle$ denotes the ground state of $H_{g}$ 
which includes all the $\sigma^{z}$-basis with 
an equal weight $1/\sqrt{2^{N}}$. 

Thanks to the specialty of the TREM~(\ref{trem}), 
we can easily say $|G(t)|^2=1$ if the quench does not go across 
the critical value $\Gamma_{\rm c}(0)$. 
On the other hand, when the quench goes across $\Gamma_{\rm c}$, 
$|G(t)|^2$ is expected to exponentially decrease as time $t$ grows, 
i.e. $|G(t)|^2\propto \exp(-Ng(t))$. 
To assess the rate function $g(t)$, Eq.~(\ref{trem}) again makes 
the problem quite simple. 
Using this, we can rewrite 
\be
 G(t) = \langle{\rm QP}|e^{-itH_{\rm REM}}|{\rm QP}\rangle
 = \frac{1}{2^N}Z_{\rm REM}(it), \label{Gtrem}
\ee
where we write the partition function of the REM under 
the inverse temperature $\beta$ as $Z_{\rm REM}(\beta)$. 
Hence, $g(t)$ is directly related to the free energy of the REM 
under the imaginary temperature.

The calculation of Eq.~(\ref{Gtrem}) is a simple task in the REM.
Using the spectral representation $G(t)=2^{-N}\sum_{n=1}^{2^N}\exp(-iE_nt)$, 
we divide the square of the amplitude $|G(t)|^2$ into 
time-dependent and -independent parts.
Taking the random average, we obtain
\be
 [|G(t)|^2] \sim \exp(-N\ln 2)+\exp(-Nt^2/2).
\ee
We naively expect that 
the first term is much smaller than the second one since 
the former involves $2^N$-terms and the latter 
$2^{N}(2^N-1)\sim 2^{2N}$.
However, when $t$ becomes large enough, the second Gaussian term 
is considerably smaller than the first term and the time-independent 
part becomes important.
Thus, we have a dynamical phase transition.
The rate function is given by 
\be
 g(t) = \left\{\begin{array}{ll} 
 \frac{t^2}{2} & \mbox{for}\ t\le t_{\rm c}=\sqrt{2\ln 2} \\
 \ln 2 & \mbox{for}\ t>t_{\rm c}
 \end{array}\right..
 \label{gt}
\ee
We see that the amplitude freezes at $t=t_{\rm c}$.


\subsection{Generalized overlap}

To obtain further information about $|\psi(t)\rangle$, 
overlaps with other wave functions are also useful. 
As such reference wave functions, we here introduce 
wave functions $|\beta\rangle$ reflecting finite temperature properties 
\be
 |\beta\rangle = 
 \frac{e^{-\beta H_{\rm REM}/2}|{\rm QP}\rangle}
 {\sqrt{\langle{\rm QP}|e^{-\beta H_{\rm REM} } |{\rm QP}\rangle}},
\ee
and consider the overlap with $|\psi(t)\rangle$
\be
 G_{1/2}(\beta,t)\equiv \langle \beta|\psi(t)\rangle
 = \frac{1}{\sqrt{2^N}}\frac{Z_{\rm REM}\left(\frac{\beta}{2}+it\right)}
 {\sqrt{Z_{\rm REM}(\beta)}}. \label{G1/2}
\ee
We can easily confirm that the quantum-mechanical average of 
physical quantities commutative with $H_{\rm REM}$ by $|\beta\rangle$ 
completely agrees with the statistical-mechanical average 
at inverse temperature $\beta$~\cite{Somma:07,Morita:08}.
On the other hand, concerning any non-commutative quantity $A$, 
the quantum-mechanical average by $|\beta\rangle$ involves non-diagonal contributions, i.e. $\sum_{n \neq m}e^{-\beta(E_n+E_m)/2}\langle {E_n}| A | E_m\rangle$, which leads to a deviation from the correct statistical-mechanical average.
However, it is known that the correct average, vanishing of 
non-diagonal contributions, can be reproduced by appropriately 
extending the Hilbert space of $|\beta\rangle$, 
the extension of which is called thermofield 
dynamics~\cite{Fano:52,Suzuki:98,Tasaki:98}. 
We can use this extension, but it does not change the following discussion 
and hence we just use $|\beta \rangle$ hereafter. 
As another way to reproduce the correct statistical-mechanical average, 
it is known that the introduction of random numbers is also 
useful~\cite{Goldstein:06,Sugiura:12}. 
This prescription is interesting but involves some technical difficulties, 
and again we do not use the prescription. 

To treat $G(t)$ and $G_{1/2}(\beta,t)$ in the same manner, 
we introduce a generalized overlap with index $k$ as 
\be
 G_k(\beta,t) = \frac{1}{(2^N)^{1-k}}
 \frac{Z_{\rm REM}\left(k\beta+it\right)}{(Z_{\rm REM}(\beta))^{k}}.
 \label{eq:G_k}
\ee
This reduces to Eq.~(\ref{G1/2}) at $k=1/2$, and $G_k(0,t)=G_0(\beta,t)=G(t)$.
Furthermore, $G_1(\beta,t)$ denotes the thermal average of 
the time evolution operator $\exp(-itH_{\rm REM})$ and 
is worth studying.

\section{Analysis and Result}

To handle the average $[|G_k|^2]$, we use the replica method 
\be
 [|G_k|^2] &=& \frac{1}{(2^{N})^{2(1-k)}}\lim_{n_1\to 1}\lim_{n_{2}\to -2k}
 \Bigl[ 
 Z_{\rm REM}^{n_1}(k\beta+it)
 \no \\ 
 & & \times Z_{\rm REM}^{n_1}(k\beta-it)Z_{\rm REM}^{n_2}(\beta)
 \Bigr].
\ee
We treat $n_1$ and $n_2$ as integers and consider the analytic 
continuation to real variables after the calculation. 
For readers familiar with the replica method, 
the introduction of $n_1$ may seem to be strange 
since the limiting value is unity, at which 
the replica symmetry breaking (RSB) effect does not appear~\cite{vanHemmen:79}. 
However, we stress that the introduction of $n_1$ is needed to construct 
the RSB solution in the numerator of Eq.~(\ref{eq:G_k}), 
which is necessary for the correct solution.
 
The calculation goes along the standard manner~\cite{Takahashi:11} and 
the result is expressed in terms of the order parameter matrix $Q$
with the size $n_1+n_1+n_2$.
We obtain $[|G_k|^2] = \lim_{n_1\to 1}\lim_{n_{2}\to -2k}\exp(-N\tilde{g})$ where
\be
 \tilde{g} &=& 
 -\frac{k^2\beta^2-t^2}{4}\sum_{a,b}^{n_1}
 \left(q_{ab}^{(+)}+q_{ab}^{(-)}\right)
 -\frac{\beta^2}{4}\sum_{a,b}^{n_2}q_{ab}^{(2)} \no \\ 
 && -\frac{ik\beta t}{2}\sum_{a,b}^{n_1}\left(q_{ab}^{(+)}-q_{ab}^{(-)}\right)
 -\frac{k^2\beta^2+t^2}{2}\sum_{a,b}^{n_1}q_{ab}^{(+-)} \no \\ 
 && -\frac{\beta(k\beta+it)}{2}\sum_{a}^{n_1}\sum_{b}^{n_2}q_{ab}^{(+2)}
 -\frac{\beta(k\beta-it)}{2}\sum_{a}^{n_1}\sum_{b}^{n_2}q_{ab}^{(-2)} \no \\ 
 && 
 +2(1-k)\ln2-s.
\ee 
$q=0$ or 1 represents a matrix element of $Q$ and 
is determined by the extremized condition.
The superscript denotes the block in the matrix $Q$: 
$+$ ($-$) is for the first (second) block with the size $n_1$, 
and 2 is for the third with $n_2$. 
The subscript denotes replica numbers in each block.
The number of configurations for a given order parameter $Q$ 
is denoted by $\exp(Ns)$ 
and $s$ plays the role of the entropy density.

Following the discussions in~\cite{Takahashi:11}, 
we can find four possible saddle-point solutions.
In the replica symmetric (RS) level, 
we have two solutions and 
the rate function $g_k= \lim_{n_1\to 1}\lim_{n_2\to -2k}\tilde{g}$ 
is calculated as follows:
\begin{description}
\item[P1]{
This solution is denoted by the identity matrix $Q=1$.
That is: $q_{ab}^{(+)}=q_{ab}^{(-)}=q_{ab}^{(2)}=\delta_{a,b}$
and zero for the other elements.
The corresponding entropy is $s=(2n_1+n_2)\ln 2$.
Then, the rate function is 
\be
 g_k=\frac{1}{2}t^2+\frac{k(1-k)}{2}\beta^2.
\ee
}
\item[P2]{
The first and second blocks have an identical configuration as 
$q_{ab}^{(+)}=q_{ab}^{(-)}=q_{ab}^{(2)}=q_{ab}^{(+-)}
=\delta_{a,b}$~\cite{Takahashi:11}.
The entropy is given by $s=(n_1+n_2)\ln 2$ and we obtain 
\be
 g_k = k\left(\frac{1}{2}-k\right)\beta^2+\ln 2.
\ee
} 
\end{description}
Other two solutions are obtained by assuming the one-step replica 
symmetry breaking (1RSB). 
In the present problem, we divide $n_1$ replicas into $n_1/m_1$ blocks 
of size $m_1$, and $q_{ab}$ takes $1$ if $a$ and $b$ are in the same block 
and $0$ otherwise. 
The $n_2$ replicas are similarly divided into $n_2/m_2$ blocks of size $m_2$. 
Two different solutions emerges according to the value of $q^{(+-)}$:
\begin{description}
\item[1RSB1]{
$q^{(+-)}=0$. 
The entropy is $s=(2n_1/m_1+n_2/m_2)\ln 2$ and the rate function becomes
\be
 g_k(m_1,m_2) &=& 2(1-k)\ln2 
 -\frac{k^2\beta^2-t^2}{2}m_1-\frac{2\ln 2}{m_1} \no\\
 & & +\frac{k\beta^2}{2}m_2+\frac{2k\ln 2}{m_2}.
\ee
Extremization with respect to $m_2$ give 
$m^*_2=2\sqrt{\ln 2}/\beta\equiv \beta_{\rm c}/\beta$. 
On the other hand, the extremization of $m_1$ gives 
a physically-unacceptable condition 
$m_1=\beta_{\rm c}/\sqrt{k^2\beta^2-t^2}$. 
Hence, we get one possible solution in this ansatz
\be
 g_k(1,m^*_2)= \frac{1}{2}t^2-\frac{1}{2}(k\beta-\beta_{\rm c})^2
 +2(1-k)\ln2.
\ee
}
\item[1RSB2]{
We assume that $q^{(+-)}$ takes the same form as $q^{(+)}$ and $q^{(-)}$.
Then, 
\be
 g_k(m_1,m_2) &=& 2(1-k)\ln2-k^2\beta^2m_1-\frac{\ln 2}{m_1} \no \\ 
 & & +\frac{k\beta^2}{2}m_2+\frac{2k\ln 2}{m_2}.
\ee 
The extremization gives $m^{*}_1=\beta_{\rm c}/2k\beta$ and 
$m^{*}_2=\beta_{\rm c}/\beta$, both of which are acceptable 
in contrast to the 1RSB1 case. 
Hence, we have three possible solutions in this case
\be
 & & g_k(1,m^{*}_2)=-\left(k\beta-\frac{\beta_{\rm c}}{2}\right)^2+2(1-k)\ln 2, \\
 & & g_k(m^{*}_1,1)= \frac{k}{2}(\beta-\beta_{\rm c})^2+2(1-k)\ln 2, \\
 & & g_k(m^{*}_1,m^{*}_2)=2(1-k)\ln 2.
\ee
}
\end{description}

\begin{figure}[htbp]
\begin{center}
\includegraphics[width=0.9\columnwidth]{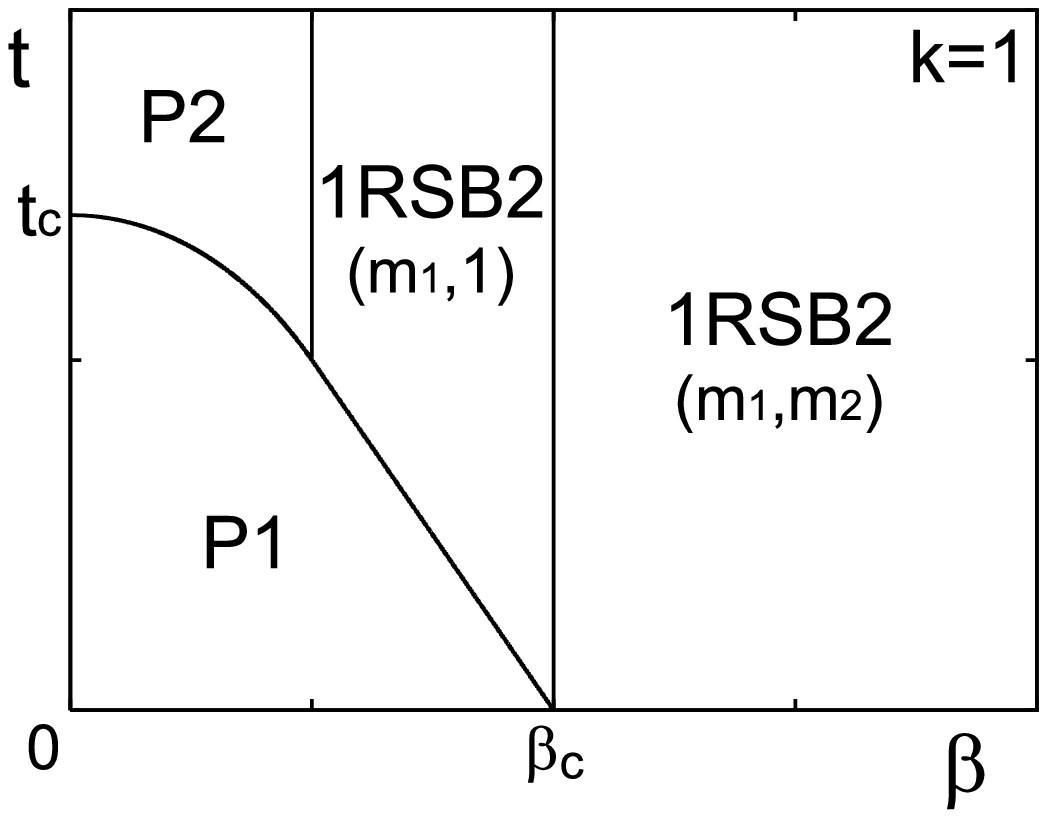}
\includegraphics[width=0.9\columnwidth]{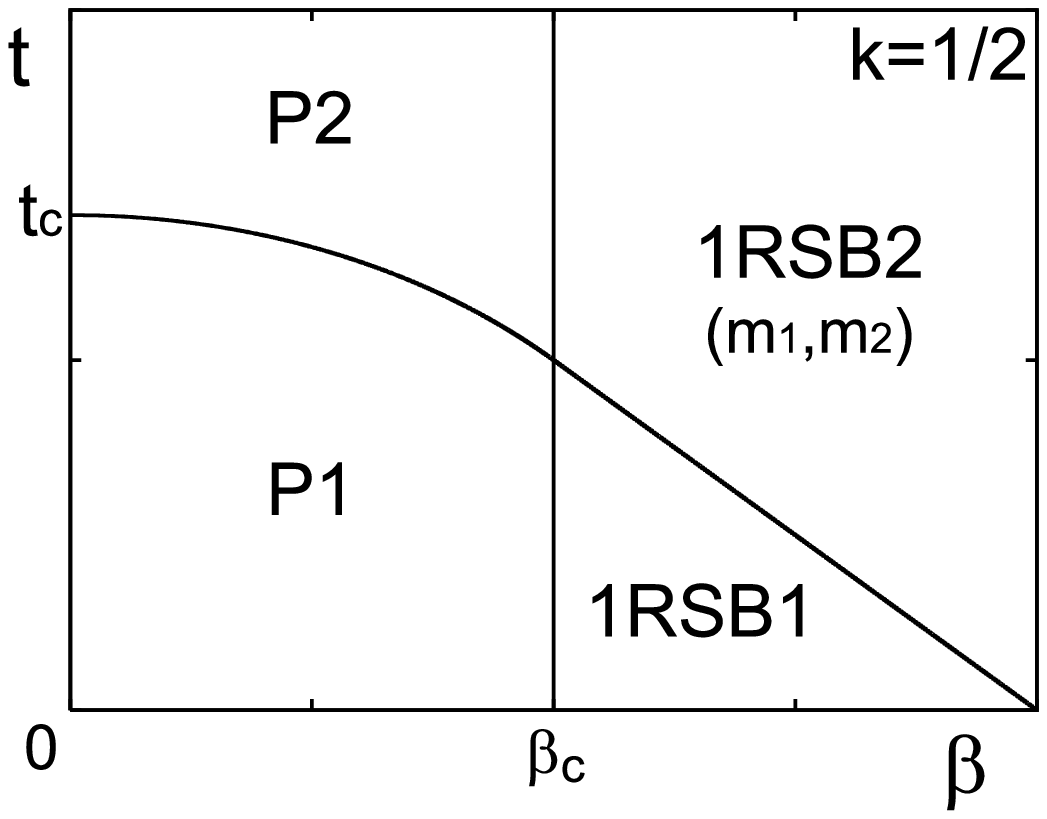}
\caption{Phase diagram for $k=1$ and $k=1/2$ on $\beta$-$t$ plane.}
\label{fig-k}
\end{center}
\end{figure}

Comparing the above solutions and considering the RSB constraint 
$0\le m_1,m_2 \le 1$, we specify the dominating solutions, 
to find that the set of solutions depend on $k$. 
For $1\le k$, the solutions are 
P1, P2, and 1RSB2 with $(m_1,m_2)=(m^{*}_1,1)$, $(m^{*}_1,m^{*}_2)$. 
For $1/2\leq k< 1$, 1RSB1 is included to the above four solutions. 
For $k < 1/2$, 1RSB2 with $(1,m^{*}_2)$ appears instead of 
that with $(m^{*}_1,1)$.
We show the phase diagram at $k=1$ and $1/2$ in Fig.~\ref{fig-k}.
In P1 and 1RSB1 phases, the rate function depends on $t$ and 
the amplitude decays in $t$.
Then, in P2 and 1RSB2 phases, the amplitude ceases to decrease and 
is frozen to a fixed value.

\section{Discussions}

\begin{figure}[htbp]
\begin{center}
\includegraphics[width=0.9\columnwidth]{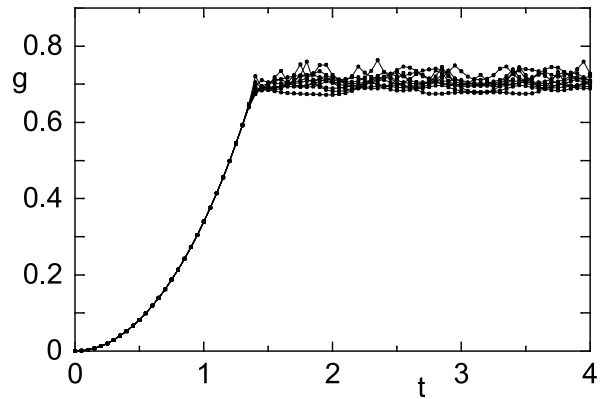}
\caption{The rate function $g(t)$ in the discrete REM 
with $N=80$ and $\alpha=1.3$.
Each line represents the result from a single sample.}
\label{fig-b0}
\end{center}
\end{figure}
\begin{figure}[htbp]
\begin{center}
\includegraphics[width=0.9\columnwidth]{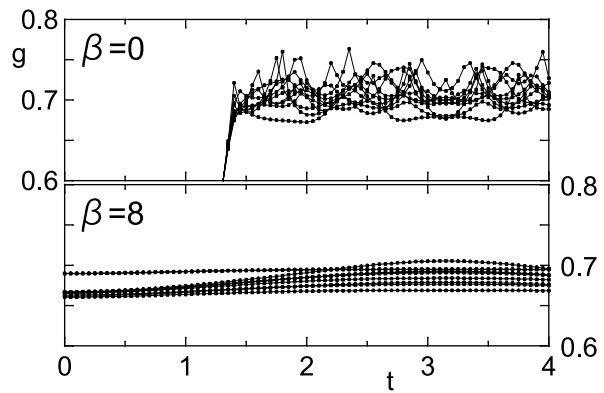}
\caption{$g_{1/2}(\beta,t)$ in the same calculation as Fig.~\ref{fig-b0}.}
\label{fig-g}
\end{center}
\end{figure}

Our result of $[|G_k|^2]$ is compared with the distribution of 
the partition-function zeros in the complex-$\beta$ plane. 
The zeros of $Z_{\rm REM}(\beta)$ with complex $\beta$ 
was calculated in Refs.~\cite{Derrida:91,Takahashi:11} and 
a similar phase diagram including the P1, P2, and 1RSB phases was found. 
In the present problem, 
the rate function $g_k$ for a given $\beta$ is frozen to 
a fixed value in the P2 and 1RSB2 phases and 
it seems to be hard to recognize the difference between these two phases.
However, it is known that the density of zeros 
two-dimensionally distributes in the P2 phase but no zeros exist 
in the 1RSB2 phase~\cite{Takahashi:11,Derrida:91}, 
which implies a qualitative difference.

Concerning the 1RSB2 phase, it is easy to interpret 
the plateau of $g_k$ with respect to $t$, 
since this corresponds to the SG phase which is well understood. 
The SG transition of the REM occurs as the freezing of the system 
into the ground state. 
This directly leads to $g_k$ in the 1RSB2 phase 
since the ground state of the REM, $|\rm SG\rangle$, 
gives the overlap $|\langle{\rm SG}|\psi(t)\rangle|^2=1/2^N$.

On the other hand, such a freezing does not occur in the P2 phase.
Instead, it is expected that the dominant states for the free energy 
rapidly switch as $t$ changes, 
which implies a kind of chaos emerges in the overlap in the P2 phase. 
This speculation is based on the recently-proposed relation 
between the distribution of zeros and the chaos effect 
of SGs~\cite{Obuchi:12}. 
The chaos effect of SGs means that the spin configuration drastically 
changes as the physical parameters, such as temperature or external field, 
slightly vary~\cite{Bray:87}. 
In the mean-field level, the chaos effect of SGs can be interpreted 
as the changes of dominant thermodynamic pure states~\cite{Rizzo:06}, 
which produces the two-dimensionally distributing zeros~\cite{Obuchi:12}. 
These considerations naturally lead to the connection 
between a kind of chaos in dynamic behavior and the P2 phase.

To confirm this assertion, we numerically calculate the rate function.
From a technical reason, we treat the discrete version of the REM, 
where $2^N$-energy levels are denoted by $M+1$-discrete 
binomially-distributed variables~\cite{Mourkarzel:91,Ogure:04}.
This model gives a phase transition for $\alpha=M/N>1$, 
whose physical nature is the same as the REM. 
We plot the rate function $g(t)$, which is equivalent to $g_{1/2}(\beta=0,t)$, 
at $\alpha=1.3$ in Fig.~\ref{fig-b0}. 
The plotted curves are taken from 10 samples of random distributions.
We can find that for small $t$, where the P1 phase dominates, 
all the samples give an identical smooth curve, 
while for large $t$ singular fluctuations depending on samples 
appear around the plateau of the P2 phase $g(t)=\ln 2$. 
This singular behavior is just the chaotic state mentioned above. 
To make further comparison, in Fig.~\ref{fig-g} 
we also plot the data at $\beta=8$ where the 1RSB2 phase dominates. 
Although the sample dependence is large both for the P2 and the 1RSB2 cases, 
there is no singular fluctuations for the latter case. 
This difference originates from the partition-function zeros, 
which are present and absent in the P2 and 1RSB2 phases, respectively.

\begin{figure}[htbp]
\begin{center}
\includegraphics[width=0.9\columnwidth]{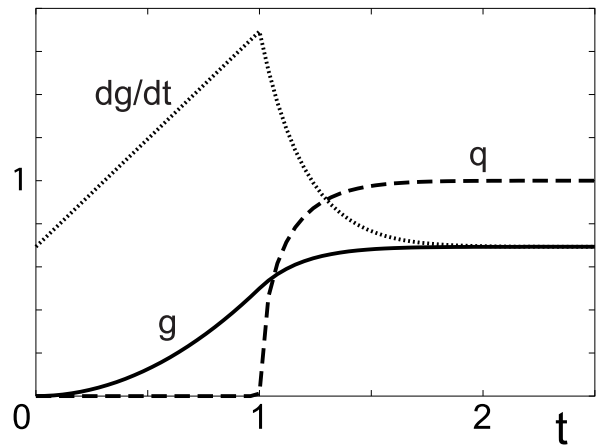}
\caption{The analysis of $[|G_{1/2}(\beta=0,t)|^2]$ in the SK model.
The solid line represents the rate function $g(t)$.
The singularity appears at $t=1$ which can be understood 
from the behavior of the order parameter $q$ (dashed line)
and the cusp of $dg(t)/dt$ (dotted line).}
\label{fig-SK}
\end{center}
\end{figure}

Note that the dynamical singularities investigated above are 
not specific to the REM. 
Actually, the presence of the P2 phase is confirmed 
in other SG models~\cite{Obuchi:12,Matsuda:10}. 
It is relatively easy to demonstrate this fact 
for the case $k=1/2$ and $\beta=0$, 
and we here show the result of the Sherrington-Kirkpatrick (SK) 
model~\cite{Sherrington:75}, which is the standard mean-field model 
of SGs and corresponds to the $p=2$ case in Eq.~(\ref{eq:SG-Hamiltonian}). 
The explicit formula of $g(t)=g_{1/2}(\beta=0,t)$ becomes
\be
 g(t)=\frac{t^2}{2}+\frac{t^2}{2}q^2-\ln \cosh \left(qt^2\right),  
\ee
where the order parameter $q$ is determined through 
\be
 q=\tanh \left(qt^2\right).
\ee
We note that there is no need to use the replica method 
in the case of $\beta=0$ since 
the denominator in Eq.~(\ref{eq:G_k}) reduces to a constant.
The order parameter $q$ represents the overlap between two spaces and 
corresponds to $q^{(+-)}$ in the analysis of the previous section.
We plot $g(t)$ in Fig.~\ref{fig-SK}, 
to find the transition of second-order at $t_{\rm c}=1$. 
The rate function $g(t)$ is more smooth than that of the REM and 
still increases $t>t_{\rm c}$, 
implying the SK model do not freeze into a small number of states 
even at $t>t_{\rm c}$, which is in contrast to the REM. 
This fact might mean that the chaotic effect more drastically affect 
the dynamical behavior in the SK model, 
which should be confirmed in future works. 

Finally, we discuss the possibility of observing 
the dynamical singularities in experiments. 
Let us consider only the case $k=1/2$ and $\beta=0$. 
What we have calculated so far is the probability that 
the time-evolved wave function becomes the state $|\psi_0 \rangle$ 
with the maximum magnetization in the $x$ direction. 
Thus, generally speaking, we can experimentally estimate this probability 
by observing $x$-direction magnetization of $|\psi(t)\rangle$ many times 
and counting the number of times where the $x$-direction magnetization 
is maximized, and hence we can see the singularity 
in the probability in principle. 
This is a kind of quantum tomography. 
Admittedly, this is quite difficult since the probability tends to be very small, 
which requires an enormous number of preparations 
and observations of the wave function. 
Thus, another pathway to observe this probability, if exists, is desired. 
Although exploration of the pathway is beyond our purpose in the present paper, 
we here address some interesting possibility. 
In equilibrium statistical mechanics, the celebrated Einstein's fluctuation theory 
tells us that the probability, or the rate function, of occurrence of atypical fluctuations 
can be directly connected to the free energy which is more easily observed in experiments. 
This connection recently inspires an exploration of similar relation in non-equilibrium phenomena, 
and some positive results are obtained~\cite{Nemoto:11,Nemoto:12}.
Similar results might be obtained for purely quantum systems, 
which can contribute to the above problem in quantum tomography. 

\section{Conclusion}

In this paper, we have studied the dynamical singularities 
of glassy systems in a quantum quench protocol, 
by treating the REM as an illustrative example. 
To widely investigate this problem, we have proposed a class of wave 
functions reflecting finite-temperature properties, 
and have defined and studied overlaps between the finite-temperature 
wave function and the time-evolved one in a quantum quench, 
not only the overlap with the initial state. 
To analyze the overlaps in a general way, we have invented a formulation 
based on the replica method and applied it to the REM. 
The rate functions of the overlaps show freezing behavior 
at critical time $t_{\rm c}$ which vanishes for low temperatures 
due to the emergence of the SG phase. 
We have also discussed the connection between the freezing behavior 
and the zeros of the partition function, 
to find a chaotic behavior for $t>t_{\rm c}$ detected by the zeros. 
The presence of this chaotic behavior is common for a wide range 
of SG models, and a demonstration on the SK case has been also presented. 

Possibility of observing the dynamical singularities in experiments 
has been also discussed. 
Although it is in general possible to observe the singularities 
in experiments, some difficulty should present due to 
the small probability of the desired event.
To resolve this problem, a better theoretical comprehension 
among non-equilibrium statistical physics and quantum many-body dynamics 
will be needed. 
There is still a gap between statistical physics 
and quantum physics communities, and we hope that this paper 
contributes to filling this gap and encouraging 
the investigation of this interdisciplinary research field.

\section*{Acknowledgment}

The authors are grateful to Y. Hashizume, Y. Kabashima, T. Nemoto, M. Ohzeki, 
and S. Sugiura for useful discussions. 
T. O. acknowledges the support by Grant-in-Aid for JSPS Fellows. 
A part of numerical calculations were carried out 
at the Yukawa Institute Computer Facility.



\begin{thebibliography}{99}

\bibitem{Onsager:44} L. Onsager, Phys. Rev. {\bf 65}, 117 (1944).

\bibitem{Yang:52} C. N. Yang and T. D. Lee, Phys. Rev. {\bf 87}, 404 (1952); 
T. D. Lee and C. N. Yang, Phys. Rev. {\bf 87}, 410 (1952).

\bibitem{Fisher:64} M. E. Fisher, 
in {\it Lectures in Theoretical Physics}, edited by W. E. Brittin 
(University of Colorado Press, Boulder, 1965), Vol. 7c.

\bibitem{Greiner:02} M. Greiner, O. Mandel, T. Esslinger, T. H\"{a}nsch, 
and I. Bloch, Nature {\bf 419}, 51 (2002).

\bibitem{Kinoshita:06} T. Kinoshita, T. Wenger, and D. Weiss, 
Nature {\bf 440}, 900 (2006).

\bibitem{Sadler:06} L. E. Sadler, J. M. Higbie, S. R. Leislie, 
M. Vengalattore, and D. M. Stamper-Kurn, Nature {\bf 443}, 312 (2006).

\bibitem{Pollmann:10} F. Pollmann, S. Mukerjee, A. G. Green, and J. E. Moore, 
Phys. Rev. E {\bf 81}, 020101 (2010).

\bibitem{Kolodrubetz:11} M. Kolodrubetz, B. K. Clark, and D. A. Huse,
Phys. Rev. Lett. {\bf 109}, 015701 (2012).

\bibitem{Heyl:12} M. Heyl, A. Polkovnikov, and S. Kehrein, arXiv:1206.2505.

\bibitem{Derrida:80} B. Derrida, Phys. Rev. Lett. {\bf 45}, 79 (1980).

\bibitem{Gross:84} D. J. Gross and M. M\'ezard, 
Nucl. Phys. B {\bf 240}, 431 (1984).

\bibitem{Goldschmidt:90} Y. Y. Goldschmidt, Phys. Rev. B {\bf 41}, 4858 (1990).

\bibitem{Obuchi:07} T. Obuchi, H. Nishimori, and D. Sherrington, 
J. Phys. Soc. Jpn. {\bf 76}, 054002 (2007).

\bibitem{Jorg:08} T. J\"{o}rg, F. Krzakala, J. Kurchan, and A. C. Maggs, 
Phys. Rev. Lett. {\bf 101}, 147204 (2008).

\bibitem{Takahashi:11} K. Takahashi, 
J. Phys. A: Math. Theor. {\bf 44}, 235001 (2011).

\bibitem{Kirkpatrick:89} T. R. Kirkpatrick, D. Thirumalai, and P. G. Wolynes, 
Phys. Rev. A {\bf 40}, 1045 (1989).

\bibitem{Bryngelson:87} J. D. Bryngelson and P. G. Wolynes, 
Proc. Natl. Acad. Sci. USA {\bf 84}, 7524 (1987).

\bibitem{Derrida:91} B. Derrida, Physica A {\bf 177}, 31 (1991).

\bibitem{Somma:07} R. D. Somma, C. D. Batista, and G. Ortiz, 
Phys. Rev. Lett. {\bf 99}, 030603 (2007).

\bibitem{Morita:08} S. Morita and H. Nishimori, 
J. Math. Phys. {\bf 49}, 125210 (2008).

\bibitem{Fano:52} U. Fano, Rev. Mod. Phys. {\bf 29}, 74 (1952).

\bibitem{Suzuki:98} M. Suzuki, J. Stat. Phys. {\bf 42}, 1047 (1998).

\bibitem{Tasaki:98} H. Tasaki, Phys. Rev. Lett. {\bf 80}, 1373 (1998).

\bibitem{Goldstein:06} S. Goldstein, J. L. Lebowitz, R. Tumulka, and N. Zangh\`i, 
Phys. Rev. Lett. {\bf 96}, 050403 (2006).

\bibitem{Sugiura:12} S. Sugiura and A. Shimizu, 
Phys. Rev. Lett. {\bf 108}, 240401 (2012).

\bibitem{vanHemmen:79} J. L. van Hemmen and R. G. Palmer, 
J. Phys. A: Math. Gen. {\bf 12}, 563 (1979).

\bibitem{Obuchi:12} T. Obuchi and K. Takahashi, 
J. Phys. A: Math. Theor. {\bf 45}, 125003 (2012). 

\bibitem{Bray:87} A. J. Bray and M. A. Moore, 
Phys. Rev. Lett. {\bf 58}, 57 (1987).

\bibitem{Rizzo:06} T. Rizzo and H. Yoshino, Phys. Rev. B {\bf 73}, 064416 (2006).

\bibitem{Mourkarzel:91} C. Moukarzel and N. Parga, Physica A {\bf 177}, 24 (1991).

\bibitem{Ogure:04} K. Ogure and Y. Kabashima, 
Prog. Theor. Phys. {\bf 111}, 661 (2004). 

\bibitem{Matsuda:10} Y. Matsuda, M. M\"uller, H. Nishimori, T. Obuchi, 
and A. Scardicchio, J. Phys. A: Math. Theor. {\bf 43}, 285002 (2010).

\bibitem{Sherrington:75} D. Sherrington and S. Kirkpatrick, 
Phys. Rev. Lett. {\bf 35}, 1792 (1975).

\bibitem{Nemoto:11} T. Nemoto and S.-I. Sasa, 
Phys. Rev. E {\bf 83}, 030105 (2011); {\it ibid.} {\bf 84}, 061113 (2011).

\bibitem{Nemoto:12} T. Nemoto, Phys. Rev. E {\bf 85}, 061124 (2012).

\end{thebibliography}
\end{document}